\documentclass[prb,twocolumn,showpacs,amssymb]{revtex4-1}

\usepackage{amsmath, amssymb}
\usepackage[dvips]{graphicx}

\begin{document}
\title{Spin polarization oscillations in a magnetically inhomogeneous conducting ring}
\author{P.V.Pyshkin}
\email{pavel.pyshkin@gmail.com}
\affiliation{B.Verkin Institute for Low Temperature Physics \&
Engineering,  National Academy of Sciences of Ukraine, 47 Lenin
Ave, Kharkov, 61103, Ukraine}

\begin{abstract}
The real part of the conductance of a 1D conducting ring with magnetic properties that vary
along the conductor is examined. The possibility of exciting a new type of spin polarization oscillations
in the ring with an external emf is analyzed. 
\end{abstract}
 \maketitle
It has been shown \cite{ref1,ref2,ref3} that new, weakly damped spin polarization
waves which may be accompanied by electric current
oscillations, can exist in micro-sized magnetic conductors
and their frequency can be controlled over a wide range
through variations in external parameters. This sort of ``spin
pendulum'' may be regarded as another spintronic device \cite{ref4}.
Combined oscillations of the spin concentration and drift
current in conducting magnetic rings have been examined \cite{ref1,ref2}
under conditions corresponding to electron hydrodynamics \cite{ref5}.
These oscillations can exist only in a ring with spatially inhomogeneous
magnetic properties: The equilibrium densities
of the spin components $n_\uparrow$ and $n_\downarrow$
(``spin-up'' and ``spindown'' electrons) must vary along the ring. Magnetic inhomogeneity
can be produced, for example, in conducting
heterostructures \cite{ref6} by means of a nonuniform electrical gate
potential.

The existence of a different type of oscillations in the
spin density under ballistic electron transport conditions in
rings and open circuits has been pointed out \cite{ref3}. These oscillations
should be most marked in one-dimensional conductors,
i.e., wires that are so thin that only one conductivity channel
undergoes spatial quantization. These oscillations are similar
to beam waves \cite{ref7} in metals with several groups of charge carriers.
Electrical neutrality during ballistic transport by electrons
with a nonequilibrium spin density $\rho_{\uparrow,\downarrow}$ (the total
charge density of the spin components is  $n_{\uparrow,\downarrow} + \rho_{\uparrow,\downarrow}$) is
achieved by meeting the condition $\rho_\uparrow = -\rho_\downarrow$. (Here and in the
following we are considering one-dimensional densities.)
These oscillations do not require magnetic inhomogeneity of
the wire; they are possible both in magnetic and nonmagnetic
materials. However, as shown below, they can be excited by
an electric field only in the magnetically nonuniform case.
As opposed to spin waves in metals \cite{ref8}, the spin polarization
oscillations predicted in \cite{ref3} exist even when there is no
spin dependent Fermi-liquid interaction.

\begin{figure}
\includegraphics[width=8cm]{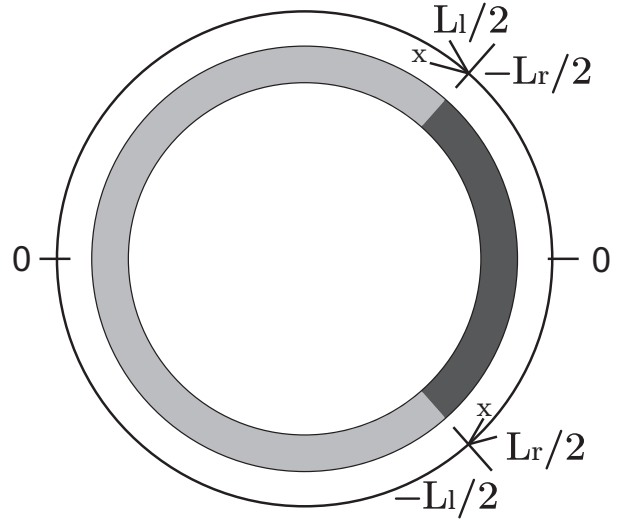}
\caption{\label{fig1} A conducting ring consisting of two parts with different magnetic
properties.} 
\end{figure}

This paper examines the excitation of all oscillations of
this type in magnetically inhomogeneous rings with an arbitrary
relation between the frequencies of collisions with and
without momentum loss and the frequency of the oscillations.
We consider a ring consisting of a 1D conductor whose
magnetic inhomogeneity is modelled as two homogeneous
magnets with lengths $L_l$ and $L_r$ Fig.\ref{fig1} with  equilibrium
densities $n_{\uparrow , \downarrow l}$ and $n_{\uparrow , \downarrow r}$, respectively, of the spin components
in the ``left'' and ``right'' parts of the ring. This model has
been chosen in order to simplify the calculations and it is not
necessary to join two different magnets in an experiment. A
contact potential barrier usually develops at a junction and
this is not present in our analysis. The directions of the magnetization
are assumed collinear at all points in the ring, so
that there are no coherence effects in the magnetic components.

The quasiclassical equations for the drift velocities of the
electron components $u_{\uparrow , \downarrow}$
and the increments in their densities
$\rho_{\uparrow , \downarrow}$ in the ballistic limit for a 1D conductor are given in
\cite{ref2}. Relaxation phenomena in a magnet can be taken into
account using the Flensberg approximation \cite{ref9}, as done in \cite{ref10}. We then have the following system of equations:
\begin{equation}
\frac{\partial\rho_\sigma}{\partial t} + j_\sigma^\prime = -\frac{e\Pi_0}{\tau_{sf}}\left(\mu_\sigma - \mu_{-\sigma}\right) \label{eq1}
\end{equation}
\begin{equation}
\frac{m}{n_\sigma}\frac{\partial j_\sigma}{\partial t} + \left(\mu_\sigma + e\varphi\right)^\prime - eE = -\sum_{\sigma'}\beta_{\sigma\sigma'}j_{\sigma'} \label{eq2}
\end{equation}
\begin{equation} \label{eq3}
\begin{split}
\beta_{\uparrow\uparrow}=e(\rho_{i\uparrow}+An_{\uparrow}^{-2}),\;
\beta_{\downarrow\downarrow}=e(\rho_{i\downarrow}+An_{\downarrow}^{-2}),\\
\beta_{\uparrow\downarrow}=\beta_{\downarrow\uparrow}=-eA(n_\uparrow
n_\downarrow)^{-1} 
\end{split}
\end{equation}
\begin{equation}
\sum_\sigma\rho_\sigma=0 \label{eq4}
\end{equation}
These equations hold for a degenerate electron system in
an approximation that is linear with respect to the deviation
from equilibrium. The subscript $\sigma$ enumerates the spin direction
($-\sigma$ is opposite to $\sigma$); $j_\sigma=n_\sigma u_\sigma$ is the electric current of the spin component; $\Pi_0^{-1} = \Pi_\uparrow^{-1} + \Pi_\downarrow^{-1}$, where $\Pi_\sigma$ is the density
of states at the Fermi level; $\tau_{sf}$ is the lifetime with respect to
spin flip of the electrons; the masses of the electrons for both
components are assumed for simplicity to be the same and
equal to $m$; $\varphi$ is the potential of the electric field that develops
with electron disequilibrium; $e$ is the electronic charge;
and, $\mu_{\sigma}$ is the nonequilibrium increment to the chemical potential,
given in the linear approximation by
\begin{equation}
\rho_\sigma =e\Pi_\sigma\mu_\sigma \label{eq5}
\end{equation}
The $\beta_{\sigma\sigma'}$ describe the electron relaxation processes: $\rho_i$ is the
specific resistance associated with momentum-loss electron
collisions (e.g., collisions with impurities) and the constant $A$
is proportional to the electron-electron collision frequency
$\nu_{ee}$ ($A\approx e^{-1}m\nu_{ee}n_m,\;n_m^{-1}=n_{\uparrow}^{-1}+n_{\downarrow}^{-1}$) with conservation of the
momentum of the electron system.

We consider the excitation of spin density oscillations in
the ring by an electric induction field  $E = E_0e^{i\omega t}$ created
by the flux of an external variable magnetic field through the
area of the ring. We limit ourselves to linear oscillations, so
in the following we assume that
\begin{gather*}
\rho_{\uparrow,\downarrow}(x,t)=\rho_{\uparrow,\downarrow}(x)e^{i\omega t}, \\
j_{\uparrow,\downarrow}(x,t)=j_{\uparrow,\downarrow}(x)e^{i\omega t}.
\end{gather*}

For further calculations it is convenient to reduce Eq. \eqref{eq1} to
the following form using Eqs. \eqref{eq4} and \eqref{eq5}:
\begin{equation}
e\Pi_\sigma\mu_\sigma\left(i\omega+\tau_{sf}^{-1}\right)+j_\sigma^\prime=0 \label{eq6}
\end{equation}
We write Eq. \eqref{eq2} as
\begin{equation}
\gamma_\sigma j_\sigma - aj = -\frac{\mu_\sigma^\prime}{e} - \varphi^\prime + E \label{eq7}
\end{equation}
where
\begin{equation*}
\gamma_\sigma = \frac{i\omega m}{en_\sigma}+\rho_{i\sigma} + \frac{A}{n_\sigma^2} + \frac{A}{n_\sigma n_{-\sigma}},\, a=\frac{A}{n_\sigma n_{-\sigma}},\, j=j_\uparrow+j_\downarrow.
\end{equation*}

Let the points $x=0$ lie in the middle of both parts, as
shown in \ref{fig1}. Then the solution of Eqs. \eqref{eq6} and \eqref{eq7} can be
conveniently sought in the form
 \begin{gather*}
 j_{\sigma}(x) = d_{\sigma} + f_{\sigma}\left(e^{-x/\lambda}+ \alpha e^{x/\lambda}\right) \\
 \mu_{\sigma}(x) = g_{\sigma}\left(e^{-x/\lambda}-\alpha e^{x/\lambda}\right)\\
 \varphi (x) = b+cx+h\left(e^{-x/\lambda}-\alpha e^{x/\lambda}\right).
 \end{gather*}

The computations are greatly simplified because of the symmetry
of the equations for $j_\sigma$, $\mu_\sigma$, $\varphi$ with respect to the substitution $x\rightarrow -x$. The symmetry means that for all solutions
(if they exist) the coefficient $\alpha$ can be chosen to have values of $\alpha =\pm 1$. If $E\neq 0$ (induced oscillations), then choosing $\alpha = -1$ does not correspond to symmetry of the induced force in Eq.
\eqref{eq2}. Thus, the way of exciting oscillations with an electric
field considered here only yields eigenmodes that are symmetric
with respect to the coordinate origin in Fig. \ref{fig1}. Setting
$\alpha  = 1$, substituting the currents and potentials in Eqs.\eqref{eq6} and \eqref{eq7}, and equating (separately) the free constants and preexponential factors, for each part of the ring we obtain the system of equations
\begin{gather*}
e\Pi_{\sigma}\left(i\omega+\tau_{sf}^{-1}\right)g_{\sigma} = \frac{f_\sigma}{\lambda}\\
\gamma_{\sigma}f_\sigma = \frac{1}{e\lambda}\left(g_{\sigma} + eh\right) \\
\gamma_{\sigma}d_{\sigma} - aj = c+E.
\end{gather*}
The condition of electrical neutrality \eqref{eq4} and the continuity
equation \eqref{eq1} imply that the total current $j$ is independent of
position, so that $f_\sigma  = -f_{-\sigma}$. The first and second of these equations
form a system of homogeneous equations for $f$, $\mu_\sigma$ and $\varphi$. This system of equations has a nontrivial solution for $\lambda$ such that
\begin{equation}
\lambda^2 = \left(e\Pi_{0}\left(i\omega+\tau_{sf}^{-1}\right)\gamma\right)^{-1} \label{eq8},
\end{equation}
where $\gamma = \gamma_{\uparrow} + \gamma_{\downarrow}$. The third of these equations can be used to express the constants $d_{\sigma}$ in terms of the total current $j$,
\begin{equation}
d_{\sigma}=j\frac{\gamma_{-\sigma}}{\gamma}. \label{eq9}
\end{equation}
The constants corresponding to the different parts of the ring
are coupled by the continuity conditions for the spin currents
and the conditions for the electrochemical potentials at the
two boundaries in the ring. Since the symmetry of the problem
is already taken into account in writing down these
quantities, it is sufficient to write down the matching conditions
at one of the two boundaries:
\begin{gather*}
j_{l\sigma}(-L_l/2) = j_{r\sigma}(L_r/2)\\
\mu_{l\sigma}(-L_l/2) + e\varphi_{l}(-L_l/2) = \mu_{r\sigma}(L_r/2) + e\varphi_{r}(L_r/2).
\end{gather*}
The above system of equations can be used to express the
constants $f_\sigma$ in terms of the constants $d_\sigma$:
\begin{equation}
f_{l,r\sigma} = \frac{(d_{r,l\sigma}-d_{l,r\sigma})(1-N_{r,l})\gamma_{r,l}\lambda_{r,l}e^{-L_{l,r}/2\lambda_{l,r}}}{(1+N_l)(1-N_r)\gamma_r\lambda_r+(1+N_r)(1-N_l)\gamma_l\lambda_l}, \label{eq10}
\end{equation}
\begin{equation*}
N_{l,r} = e^{-L_{l,r}/\lambda_{l,r}}. \notag
\end{equation*}
Substituting Eq. \eqref{eq10} in Eq. \eqref{eq7}, using Eq. \eqref{eq9} for this, and
integrating the resulting equation over the entire ring, we
obtain the relationship between the total current $j$ and the
external emf $\oint Edx$:
\begin{gather}
\oint E dx = R j \label{eq11} \\
R = R_l + R_r + 2R_{border} \label{eq12} \\
R_{l,r} = L_{l,r}\left(\frac{\gamma_{\uparrow\,l,r}\gamma_{\downarrow\,l,r}}{\gamma_{l,r}} - a\right) \label{eq13}\\
R_{border} = \frac{(1-N_l)(1-N_r)(\gamma_{\downarrow\,l}\gamma_{\uparrow\,r} - \gamma_{\downarrow\,r}\gamma_{\uparrow\,l})^2}{\gamma_l^2\gamma_r^2\left(\displaystyle{\frac{(1+N_l)(1-N_r)}{\lambda_l\gamma_l}+\frac{(1+N_r)(1-N_l)}{\lambda_r\gamma_r}}\right)}. \label{eq14}
\end{gather}

Equation \eqref{eq13} gives the resistances of the two parts of the
ring in the absence of a spin disequilibrium; thus, $R_{border}$ is
related to the magnetic inhomogeneity of the ring. If we set
$\omega=0$, $N_{l,r}=0$, then Eq. \eqref{eq14} is exactly equal to the nonequilibrium
resistance of a single boundary between two
wires with different magnetic properties obtained in Ref. \cite{ref11}.
In the following we shall be interested in the dependence of
the real part of the conductance of the ring on the frequency
$\omega$ of the external emf i.e., $Z(\omega)=\mathrm{Re}(1/R(\omega))$.

\begin{figure}[t] 
\includegraphics[width=8cm]{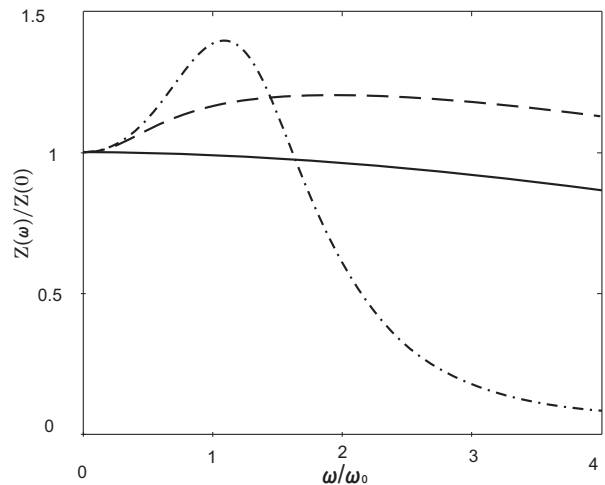}
\caption{ \label{fig2} The function $Z(\omega)/Z(0)$ constructed using Eq. \eqref{eq12} for a ring consisting of two parts of equal length ($L_l = L_r$). The frequency is measured in
units of the characteristic frequency $\omega_0^2 = \displaystyle{\frac{n_\uparrow n_\downarrow}{mne\Pi_0 L^2}}\approx\displaystyle\frac{v_F^2}{L^2}$ of the system.
The smooth and dashed curves correspond to the case where the frequencies
of normal collisions and collisions with momentum loss of the
electron system are 10 times the characteristic frequency ($\nu_{ee}=\nu_{i}=10\omega_0$).
The smooth curve is for a magnetically homogeneous ring ($n_{l\uparrow\downarrow}=n_{r\uparrow\downarrow}$) and
the dashed curve, for a magnetically inhomogeneous ring ($n_{l\uparrow\downarrow}=n_{r\downarrow\uparrow}$) with
polarization $n_\uparrow/n_\downarrow = 30/70$ of the electron density. The dot-dashed curve corresponds
to the case of a magnetically inhomogeneous ring ($n_{l\uparrow\downarrow}=n_{r\downarrow\uparrow}$) with
polarization $n_\uparrow/n_\downarrow = 30/70$ of the electron density and an electron-electron
collision frequency a factor of 10 times higher than the frequency of collisions
with momentum loss by the electron system ($\nu_{ee}=10\omega_0$, $\nu_i=\omega_0$). In all
cases the spin flip frequency is $\tau_{sf}^{-1}=\omega_0/10$}
\end{figure}

\begin{figure}[t] 
\includegraphics[width=8cm]{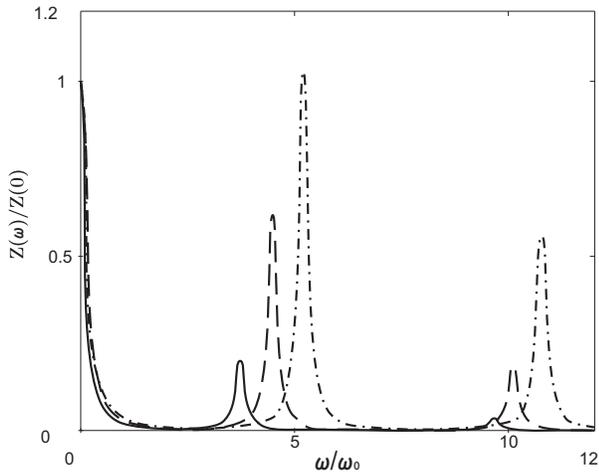}
\caption{\label{fig3} The function $Z(\omega)/Z(0)$ constructed using Eq. \eqref{eq12} for a ring consisting of two parts of equal length ($L_l = L_r$). The frequency is measured in
units of the characteristic frequency $\omega_0^2 = \displaystyle{\frac{n_\uparrow n_\downarrow}{mne\Pi_0 L^2}}\approx\displaystyle\frac{v_F^2}{L^2}$ of the system,
and the polarization $n_\uparrow/n_\downarrow$
of the electron density is $20/80$ for the
smooth curve, $10/90$ for the dashed curve, an $5/95$ for the dot-dashed curve.
The frequencies of normal collisions, collisions with momentum loss by the
electronic system, and spin flip are each a factor of 10 times smaller than the
characteristic frequency of the system, i.e., ($\nu_{ee}=\nu_{i}=\tau_{sf}^{-1}=\omega_0/10$).}

\end{figure}

\begin{figure}[t]
\includegraphics[width=8cm]{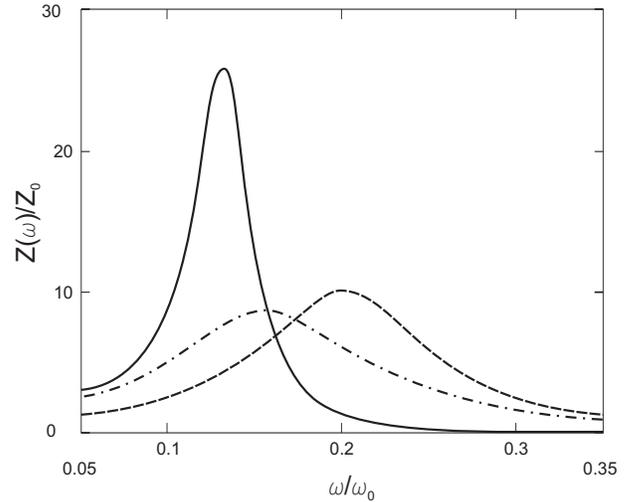}
\caption{\label{fig4} The function $Z(\omega)/Z_0$ constructed using Eq. \eqref{eq18} for a magnetically
inhomogeneous ring with a smooth change in its magnetic properties. The
frequency of normal collisions is $\nu_{ee}=1000\omega_0$ and there are no spin-flip processes
or collisions with loss of momentum by the electron system. $Z_0^{-1} = \displaystyle{\oint\frac{m\omega_0}{ne^2}dx}$. $L_{tr}=0.2L$ for the smooth curve, $L_{tr}=0.05L$ for the dashed
curve, and $L_{tr}=0.0125L$ for the dot-dashed curve, where $L$ is the length of
the ring and $L_{tr}$ is the characteristic size of the magnetic inhomogeneity.
}

\end{figure}

Figure \ref{fig2} shows that a magnetic inhomogeneity of the
ring makes $Z(\omega)$ increase with rising frequency for small $\omega$,
while $Z(\omega)$ for a magnetically homogeneous ring decreases
monotonically as the external frequency is raised. The increased
conductance at low frequencies for a magnetically
inhomogeneous sample can be explained in terms of the results
of Ref. \cite{ref11}: since a spin inhomogeneity leads to increased
electrical resistance in static conductivity \cite{ref11}, a frequency
increase that limits the disequilibrium region in
accordance with Eq. \eqref{eq8}, will increase the conductance. The
subsequent reduction in the conductance with increasing frequency
is explained (as in the case of a uniform ring) by the
inertia of the charge carriers, in accordance with the Drude-Lorentz formula. This rise in the conductance at frequencies
that are low compared to $\omega_0$ occurs when $\nu_i + \nu_{ee} > \omega_0 >> \tau_{sf}^{-1}$.

The characteristic maximum in Fig. \ref{fig2} for a hydrodynamic
conductivity corresponds to a ``spin pendulum'' resonance
near the characteristic frequency \cite{ref1}. However, as opposed
to Refs. \cite{ref1,ref2,ref3}, where a smooth variation in the magnetic
properties of the ring was assumed, when there are sharp
boundaries between regions with different magnetic properties
the eigenmodes of the ``spin pendulum'' will be highly
damped owing to diffusive processes near the boundaries and
the damping constant will be on the order of the resonance
frequency itself.

It should also be noted that the different mean free paths
for collisions with impurities by the spin groups in a
magnet \cite{ref12} can also affect the character of the resonance
peaks. Thus, if the electron spin group with the lowest density
scatters on impurities more often than electrons from
another group, then the resonance peaks will be more distinct
than when the scattering by electrons from the different spin
groups is the same.

In the limit of ballistic transport in the ring, when the
mean free path of the electrons for any collisions is much
greater than the length of the ring, i.e., $\nu_{ee},\nu_i<<v_F/L$, and the
spin relaxation time is much longer than the characteristic
time for passage of an electron around the ring, i.e., $\tau_{sf}>>L/v_F$, the behavior of the conductance of a magnetically
inhomogeneous ring differs substantially from the diffusive
or hydrodynamic cases. Instead of a single resonance maximum
in $Z(\omega)$, there are a number of resonances (Fig. \ref{fig3})
which are more distinct when the relaxation frequencies are
lower. As in the case of diffusive transport, the resonances
show up more distinctly if the magnetic inhomogeneity of
the ring is greater. For the case in which both parts of the
ring have the same length, $L_l=L_r$, the ``left'' and ``right'' parts
are oppositely polarized ($n_{l\uparrow\downarrow}=n_{r\downarrow\uparrow}$), and are also weak, i.e.,
$|n_\uparrow - n_\downarrow|/n<<1$,  with $\Pi_0=Const$, it is possible to find the resonance frequencies $\omega$ analytically. To do this, we set $\rho_{i\uparrow\downarrow}=0$
and $A=0$ in Eqs. \eqref{eq13} and \eqref{eq14}, and then solve the equation $R(\omega)=0$. As a result, we obtain the following expression for
the resonance frequencies:
\begin{gather}\label{ballist_omega}
\begin{split}
\omega = \omega_0\left(1-\frac{\delta}{8}\right)\left(\pi+2\pi k +\frac{2\delta}{\pi+2\pi k}\right),\\
\delta=4\frac{(n_\uparrow - n_\downarrow)^2}{n^2}<<1 
\end{split}
\end{gather}
\begin{equation*}
\omega_0 = \frac{1}{2}\sqrt{\frac{n}{me\Pi_0L^2}},\,k=0,1,2\dots
\end{equation*}
As can be seen from Eq. \eqref{eq14}, $R_{border} = 0$ at all frequencies
if the ratio  $\gamma_\uparrow /\gamma_\downarrow$
is independent of position, i.e., under
this condition all the above described features of the conductance
are absent. This situation occurs with a homogeneous
ring, but also if the inhomogeneity is not ``magnetic'', i.e., if
the densities of the spin components and their specific conductivities
$\rho_i$ vary from point to point in proportion to one
another. Thus, we have proved that oscillations can be excited
by an electric field in a magnetically inhomogeneous
ring. It is in just this case that spin density waves are accompanied
by a variable electric current along the ring. The relationship
between the spin oscillations and the current can
be explained qualitatively as follows: the electrical neutrality
condition makes the inhomogeneous concentrations of the
two different spin components move together along the wire.
When the conductivity of the components is nonuniform this
is achieved through the appearance in the wire of an internal
electric field, which creates a current.

It should be kept in mind that the roots of the equation
$R(\omega) = 0$ do not determine the frequencies of all the eigenmodes
of the system, but only of those which can be excited
by this electrical mechanism. Oscillations with current that
are odd in x were mentioned above; they are not excited by
an electric field, but are solutions of the system of Eqs. \eqref{eq4} - \eqref{eq7} without collisions and with $E=0$. The frequencies of
these eigenmodes are
\begin{equation*}
\omega = \omega_0\left(1-\frac{\delta}{8}\right)\left(2\pi k +\frac{\delta}{\pi k}\right).
\end{equation*}
Oscillations that are even in $x$ and odd in $x$ also exist in the
ballistic regime in a homogeneous ring ($\delta = 0$).

Since a sharp transition between magnetically homogeneous
parts of the ring leads to rapid damping of ``spin pendulum''
waves, it is clear that a smooth transition leads to the
most important changes in the case of hydrodynamic transport.
In order to follow these changes, we introduce a finite
transition length $L_{tr} << L$ and let $\nu_{ee} >> \omega_0$, while neglecting
collisions that do not conserve momentum and spin. We distinguish
a purely hydrodynamic part of the currents which
corresponds to equality of the drift velocities of the components:
\begin{equation*}
j_\sigma = j\frac{n_\sigma}{n} \pm j_1,
\end{equation*}
where the ``+'' sign corresponds to spin-up and ``-'' to spindown.
Subtracting Eq. \eqref{eq7} for one component from the other
for the different components, solving the resulting equation
for $j_1$, and substituting in Eq. \eqref{eq1}, we obtain
\begin{equation}
\frac{\partial\rho_\uparrow}{\partial t} = \left[\frac{n_0}{A}\left(\frac{\rho_\uparrow}{\Pi_0}\right)^\prime\right]^\prime - j\left(\frac{n_\uparrow}{n}\right)^\prime, \label{eq16}
\end{equation}
where $n_0^{-1} = n_\uparrow^{-1} + n_\downarrow^{-1}$. Equation \eqref{eq16} describes the diffusion of
the nonequilibrium spins near the transition, where the flowing
current generates a spin disequilibrium. We simplify the
model by assuming that $n_0$, $\Pi_0$ and $A$ are independent of $x$.
Then we have
\begin{equation}
\mu_\sigma(x) = - \frac{j}{\Pi_\sigma}\int G(x-y)\left(\frac{n_\uparrow(y)}{n(y)}\right)^\prime dy, \label{eq17}
\end{equation}
where the Green function for the diffusion equation \eqref{eq16} is
\begin{equation*}
G(x) = \frac{1}{2i\omega\lambda} e^{-|x|/\lambda}, \, \lambda = \sqrt{\frac{n_0^2}{i\omega A\Pi_0} }.
\end{equation*}

In order to obtain an expression for the resistance $R$ in this
case, we multiply both parts of Eqs. \eqref{eq7} by $n_\sigma$, add the equations
for the different components (the terms with $A$ drops
out of this sum), divide by $n$, and integrate over the entire
ring. After integration by parts and using Eq. \eqref{eq17}, we have
\begin{multline}\label{eq18}
R=\frac{1}{e^2}\oint \left[ \frac{i\omega m}{n} - \right. \\
 \left. \frac{1}{\Pi_0}\int \left(\frac{n_\uparrow(x)}{n(x)}\right)^\prime G(x-y)\left(\frac{n_\uparrow(y)}{n(y)}\right)^\prime dy\right]dx. 
\end{multline}

When $\lambda << L_{tr}$ the function $G(x - y)$ in Eq. \eqref{eq18} can be replaced
by $(i\omega)^{-1}\delta(x - y)$ and the condition $R=0$ then gives
the frequency of the oscillations of the ``spin pendulum''
found in Ref. \cite{ref1}:
\begin{equation}
\omega^2 =
{\displaystyle\oint{\left(\frac{n_\uparrow}{n}\right)'}^2
\frac{1}{\Pi_0}dx}\left(\displaystyle\oint\frac{m}{n}dx \label{eq19}
\right)^{-1}.
\end{equation}

Figure \ref{fig4} is a plot of the function $Z(\omega)$ for a ring consisting
of two oppositely magnetized parts of equal lengths with
smooth transition regions of length $L_{tr}$ between them, constructed
using Eq. \eqref{eq18} for three different values of $L_{tr}$. 
The graphs were constructed assuming weak magnetization of the ring 
($| n_\uparrow-n_\downarrow | / n=0.035$) 
so that $\lambda$
could be treated as independent of $x$. It can be seen that the most distinct conductance peak occurs with the smoothest variation in the magnetic
properties of the ring. This is explained by the fact that
the period of the eigenmodes in this case is shorter than the
time for the spin disequilibrium to diffuse through the region
of $L_{tr}$. When the length of the transition between the halves
of the rings with their different magnetic properties is reduced,
the frequency of the resonance initially increases in
accordance with the formula for the eigenfrequency \eqref{eq19} of
the ``spin pendulum'' and ceases to increase when the ``spin
pendulum'' frequency approaches $\tau_d^{-1}$, where $\tau_d \approx L_{tr}^2/D$ is the
diffusion time for the spin disequilibrium in the region $L_{tr}$
and $D \approx v_F^2/\nu_{ee} \approx \omega_0^2 L^2/\nu_{ee}$ is the diffusion coefficient.

We now estimate the characteristic order of magnitude
of these oscillations in the ballistic frequency regime, $\omega_0 \approx v_F/L$. Since the mean free path for collisions with structural
imperfections in AsGa heterostructures is as long as~10~$\mu$ and $v_F \approx 10^7$ cm/s, we have $\omega_0 \approx 10^{10}s^{-1}$. The frequency
of the oscillations of the ``spin pendulum'' (hydrodynamic
regime) can be significantly lower than $\omega_0$, since it
contains the degree of inhomogeneity, $(n_\uparrow/n)_r - (n_\uparrow/n)_l$, as a
reducing factor.

In this paper we have examined the frequency dependence
of the resistance of magnetically inhomogeneous
closed conductors. It has been shown that, depending on how
the real part of the conductance of the ring depends on the
frequency of an external emf, there can be a single maximum
(in the case of diffusive or hydrodynamic transport) or a
number of sharp peaks (in the case of ballistic transport)
corresponding to excitation of eigenmodes of the spin polarization
in the conductor by the external field. In the hydrodynamic
case, the most distinct conductance maximum can
be observed in rings with smoothly varying magnetic properties
(``spin pendulum'').

The author thanks A. I. Kopeliovich for continuing interest
in this work and valuable comments.

Translated by D. H. McNeill.

\end{document}